\documentclass[preprint,12pt]{elsarticle}

\usepackage[a4paper, margin=1in]{geometry}  

\usepackage{amssymb}
\usepackage{amsmath}
\usepackage{microtype}

\usepackage{hyperref}



\journal{Nuclear Physics B}

\begin{document}

\begin{frontmatter}



\title{Investigation of Thermodynamic Properties of Classical Oscillators Under Statistical and Superstatistical Frameworks}




\author[label1]{Huilin Wang\fnref{1}\fnref{*}}
\ead{wanghuilin@mails.ccnu.edu.cn}

\author[label1]{Weibing Deng}

\author[label1]{Zhekai Chen}

\affiliation[label1]{%
  organization={Key Laboratory of Quark and Lepton Physics (MOE) and Institute of Particle Physics, Central China Normal University},
  addressline={Wuhan 430079},
  city={Wuhan},
  postcode={430079},
  country={China}
}

\fntext[*]{Correspondence: \href{mailto:wanghuilin@mails.ccnu.edu.cn}{wanghuilin@mails.ccnu.edu.cn}}

\begin{abstract}

This paper systematically investigates the thermodynamic properties of classical oscillators under different statistical distributions, focusing on the behavior of uniform distribution, two-level distribution, gamma distribution, log-normal distribution, and F-distribution as the nonequilibrium parameter \( q \) varies. By calculating Helmholtz free energy and entropy, we reveal the unique patterns and characteristics exhibited by each distribution during the process of moving away from equilibrium. The results show that uniform and two-level distributions exhibit consistent trends of decreasing free energy and increasing entropy under nonequilibrium statistics, reflecting an increase in system disorder. In contrast, the gamma, log-normal, and F-distributions display complex dual-equilibrium point phenomena, where the system can briefly return to equilibrium at specific \( q \) values. However, as \( q \) further increases, the system rapidly moves away from equilibrium, exhibiting pronounced nonequilibrium characteristics. These findings not only deepen our understanding of nonequilibrium statistical physics but also provide new theoretical perspectives and methods for studying the nonequilibrium behavior of complex systems.
\end{abstract}




\end{frontmatter}


\section{Introduction}

The classical harmonic oscillator is a cornerstone of physics, serving as a fundamental model to understand a wide range of phenomena across disciplines. Its significance is deeply rooted in both classical mechanics and quantum mechanics, providing insights into systems ranging from atomic vibrations to macroscopic mechanical devices. This work aims to capture the essence of the harmonic oscillator, highlighting its theoretical foundations, experimental validations, and its evolving role in modern physics, all supported by extensive references.

Regarding noise interaction mechanisms, the study first distinguishes the differential effects of additive and multiplicative noise on phase space evolution of oscillators through analytical solutions of generalized Langevin equations and numerical simulations\cite{gitterman2005classical}. Particularly, when the system resides in a double-well potential, the transition rate asymmetry induced by multiplicative noise can generate remarkable stochastic resonance phenomena. The resonance peak position demonstrates a specific power-law dependence on noise intensity, offering theoretical guidance for noise engineering control. When extending to electromagnetic field coupling effects, the stochastic diffusion process of charged oscillators in constant magnetic fields exhibits anisotropic characteristics\cite{jimenez2008brownian}: the transverse diffusion coefficient exhibits significant attenuation with increasing magnetic field strength, while longitudinal diffusion behavior displays modified properties distinct from classical transport theory. These findings establish new analytical models for confined transport problems in plasma physics.

To deeply uncover the fundamental connections between classical stochastic dynamics and quantum phenomena\cite{giannakis2021quantum,dekker1981classical,bloch2013introduction,li2025multi,onah2023quadratic}, the research team developed an algebraic modeling approach\cite{alves2023algebraic}. By transforming Hamiltonian canonical equations into symplectic matrix representations and employing canonical transformations, the geometric manifold structure in phase space was revealed. Notably, the formal isomorphism between Poisson brackets and quantum commutators in oscillator systems suggests that the quantization process essentially corresponds to the probability amplitude representation of classical stochastic paths in Hilbert space. This theoretical breakthrough was validated through non-equilibrium fluctuation studies on two-dimensional electromagnetic field-coupled oscillators\cite{jimenez2009fluctuation}. The work distribution functions obtained by solving Smoluchowski equations not only strictly satisfy core fluctuation theorems of non-equilibrium thermodynamics but also demonstrate fluctuation spectra of entropy production rates that mirror statistical characteristics of quantum tunneling effects.

Particularly illuminating are the pulse excitation experiments in vacuum fluctuation fields. Numerical simulations reveal that classical oscillators under random electromagnetic vacuum backgrounds exhibit discrete excitation spectra\cite{huang2015discrete}, with spectral line intervals precisely matching the energy quantization rules of quantum harmonic oscillators. This discovery provides critical evidence supporting Stochastic Electrodynamics (SED) theory, suggesting that quantum phenomena may originate from nonlinear interactions between classical systems and stochastic background fields. Through constructing a multi-scale "noise-field-fluctuation" analytical framework, this series of studies establishes a unified descriptive paradigm between classical stochastic dynamics and quantum mechanics in harmonic oscillator systems, opening new pathways for exploring quantum behavior emergence in macroscopic classical systems.

This paper aims to review the theoretical foundations of the one-dimensional classical harmonic oscillator,explore its applications across various domains, and present the latest research advancements. By analyzing the analytical solutions, statistical properties, and nonlinear extensions of the classical harmonic oscillator, we hope to offer valuable insights for researchers in related fields.
The Hamiltonian $H(q_i, p_i)$ of $N$ classical oscillators describes the total energy of the system, consisting of potential and kinetic energy:
\begin{equation}\label{1}
 H({q}_{i},{p}_{i})=\frac{1}{2}m{\omega }^{2}{q}_{i}^{2}+\frac{1}{2m}{p}_{i}^{2}  
\end{equation}
where $q_i$ is the displacement of the $i$-th degree of freedom, $p_i$ is the momentum of the $i$-th degree of freedom, $m$ is the mass, and $\omega$ is the angular frequency. The first term represents the potential energy, proportional to the square of the displacement; the second term represents the kinetic energy, proportional to the square of the momentum. The subscript $i$ denotes the $i$-th degree of freedom in the system, applicable to multi-degree-of-freedom systems. In this paper, we apply statistical and superstatistical\cite{sattin2018superstatistics,davis2025superstatistics,metzler2020superstatistics,ourabah2024superstatistics} methods to investigate the distribution functions of classical oscillators. We study various distribution functions, such as uniform distribution, two-level distribution, gamma distribution, log-normal distribution\cite{dos2020log}, and F-distribution. We also analyze the behavior of classical oscillators under statistical and superstatistical methods and compare the results\cite{beck2003superstatistics,beck2004superstatistics,cohen2004superstatistics,abe2007superstatistics,davis2025superstatistics}. We consider the Tsallis statistics for superstatistical systems. Tsallis statistics generalize the usual Boltzmann–Gibbs statistics to measure the disorder or uncertainty of a system\cite{beck2009recent,beck2004superstatistics}. Tsallis statistics describe systems in nonequilibrium states or those with long-range interactions or memory effects. Tsallis statistics have been used to study dark energy, the mysterious force driving the accelerated expansion of the universe\cite{gravanis2020physical}. One method to study dark energy is through the holographic principle, which relates the entropy of a system to its boundary area. By using Tsallis entropy instead of Boltzmann–Gibbs entropy\cite{dos2023entropic,sattin2004superstatistics,souza2003stability}, holographic dark energy models with different properties and behaviors can be obtained. Some of these models explain current cosmological observations better than others. Tsallis statistics and holographic dark energy remain active research areas in physics and cosmology.

\section{Overview of Standard and Superstatistical Methods}

To obtain the statistical properties of classical oscillators, we require the partition function. In standard statistical mechanics, the following relation holds:
\begin{equation}
  Z=\sum_{n=0}^{\infty}\ g(E) \exp(-\beta E_{n})=\int_0^\infty \ g(E)\exp(-\beta E)\,dE,
\end{equation}
where $E$ is the energy associated with each microstate of the system, $\beta \equiv \frac{1}{K_{b} T}$ is the thermodynamic beta, $K_{b}$ is the Boltzmann constant, and $T$ is the temperature. The partition function in superstatistics differs. We first define the generalized Boltzmann factor as:
\begin{equation}
  B(E)=\int_0^\infty f(\beta) \exp(-\beta E) \,d\beta,
\end{equation}
where $f(\beta)$ is the probability density function of $\beta$. Using superstatistics, the partition function becomes:
\begin{equation}
  Z=\int_0^\infty B(E)g(E)\,dE=\int_{0}^{\infty}g(E)\,dE \int_{0}^{\infty }d\beta f(\beta ){e}^{-\beta E}
\end{equation}

These relationships originate from The partition function, first introduced by Boltzmann, can be expressed in terms of $\beta$ or $T$. Other statistical formulations were introduced by Gibbs, Einstein, Boltzmann-Gibbs, and Tsallis, leading ultimately to the superstatistical representation initially proposed by Wilk and Włodarczyk and later reformulated by Beck and Cohen.

Superstatistics addresses nonequilibrium systems with complex dynamics and large fluctuations of intensive quantities (e.g., inverse temperature, chemical potential, or energy dissipation) over long time scales. The name "superstatistics" arises from the superposition of two statistics: the statistics of $\beta$ and the standard Boltzmann factor $\exp(-\beta E)$. One statistic is associated with $\beta$, which is approximately constant within spatial regions, i.e., the standard Boltzmann factor $\exp(-\beta E)$, where $E$ is the energy in each spatial region. The other statistic describes the long-time average of fluctuations in $\beta$. In superstatistics, we define the Tsallis parameter $q$. When $q = 1$, the generalized Boltzmann factor reduces to the standard Boltzmann factor.

We can distinguish two types of superstatistics: Tsallis and Kaniadakis. In this paper, we investigate the thermodynamic properties of classical oscillators under Tsallis superstatistics and compare them with standard statistical mechanics. For the mean values of $\beta$ and $\beta^{2}$, we can write:
\begin{equation}
  \langle\beta\rangle= \int_0^\infty \beta f(\beta) \,d\beta,
\end{equation}
\begin{equation}
  \langle\beta^{2}\rangle= \int_0^\infty \beta^{2} f(\beta) \,d\beta,
\end{equation}
where $\langle\beta^{2}\rangle=\beta_{0}^{2}$, and the variance is given by:
\begin{equation}
  \sigma^2=\langle\beta^{2}\rangle - \langle\beta\rangle^{2}.
\end{equation}
In this section, we introduce several fundamental probability distributions and explore their interrelationships\cite{sadeghi2024investigation}. The uniform distribution is a symmetric distribution where all outcomes within a specified range are equally likely, making it ideal for modeling scenarios such as selecting a random point on a number line or assigning equal probabilities to different time intervals. The two-level distribution applies to systems with two distinct states of equal probability, such as a quantum particle existing in one of two energy levels or a binary switch being either on or off. The gamma distribution, characterized by its shape and scale parameters, is widely used to model waiting times in queuing systems or the distribution of insurance claim sizes. The log-normal distribution is suitable for phenomena where the variable is a product of many independent factors, such as modeling the distribution of biological cell sizes or the concentration of pollutants in the atmosphere. Lastly, the F-distribution, derived from the ratio of two scaled chi-square distributions, plays a critical role in hypothesis testing, particularly in analysis of variance (ANOVA) and regression models. By examining the relationships among these distributions, we derive their thermodynamic properties and provide a comparative analysis through graphical representations, thereby offering a comprehensive understanding of their applicability in various scientific contexts.
\subsection{Uniform Distribution}
We first discuss the uniform distribution of $\beta$, a simple model. Its distribution function over the range $0 \leq a \leq \beta \leq a+b$ is given by:
\begin{equation}
  f(\beta) = \frac{1}{b}.
\end{equation}
For the mean and variance of $\beta$, we have:
\begin{eqnarray} \nonumber
\langle\beta\rangle &=& a+\frac{b}{2} \\
\sigma^{2} &=& \frac{b^{2}}{12}.
\end{eqnarray}

\subsection{Two-Level Distribution}
This distribution model describes the case where subsystems alternate between two distinct discrete values with equal probability. The two-level distribution function is given as:
\begin{equation}
  f(\beta) = \frac{\delta(a)}{2}+\frac{\delta(a+b)}{2}.
\end{equation}
Additionally, the mean and variance are:
\begin{eqnarray} \nonumber
\langle\beta\rangle &=& a+\frac{b}{2} \\
\sigma^{2} &=& \frac{b^{2}}{4}.
\end{eqnarray}

\subsection{Gamma Distribution}
The gamma distribution of the inverse temperature $\beta$ leads to Tsallis statistics, one of the most notable examples in superstatistics. The distribution function is:
\begin{equation}
  f(\beta) = \frac{1}{b\Gamma(c)}\left(\frac{\beta}{b}\right)^{c-1} e^{-\frac{\beta}{b}},
\end{equation}
where $c > 0$ and $b > 0$. The mean and variance are given by:
\begin{eqnarray} \nonumber
\langle\beta\rangle &=& bc \\
\sigma^{2} &=& b^{2} c.
\end{eqnarray}

\subsection{Log-Normal Distribution}
The corresponding distribution function is given by:
\begin{equation}
  f(\beta)= \frac{1}{\beta s \sqrt{2 \pi}} \exp\left(-\frac{(\ln(\frac{\beta}{m}))^{2}}{2 s^{2}}\right),
\end{equation}
where $m$ and $s$ are the free parameters corresponding to the log-normal distribution. The mean and variance of $\beta$ are defined as:
\begin{eqnarray} \nonumber
\langle\beta\rangle &=& m\sqrt{\omega} \\
\sigma^{2} &=& m^{2} \omega (\omega-1).
\end{eqnarray}
The quantity $ \omega $ equals the exponent of the square of $ s $, i.e., $ \omega = e^{s^{2}} $.

\subsection{F-Distribution}
The distribution function for $ \beta \in [0,\infty] $ is given as:
\begin{equation}
  f(\beta)= \frac{\Gamma\left(\frac{\nu + \omega}{2}\right)}{\Gamma\left(\frac{\nu}{2}\right)\Gamma\left(\frac{\omega}{2}\right)} \left(\frac{b\nu}{\omega}\right)^{\frac{\nu}{2}} \frac{\beta^{\frac{\nu}{2}-1}}{\left(1+\frac{\nu \beta b}{\omega}\right)^{\frac{\nu+\omega}{2}}},
\end{equation}
where $ \nu $, $ \omega $, and $b$ are positive integer parameters. The mean and variance of $\beta$ are calculated as:
\begin{eqnarray}  \nonumber
\langle\beta\rangle &=& \frac{\omega}{b(\omega-2)} \\
\sigma^{2} &=& \frac{2 \omega^{2} (\nu+\omega-2)}{b^{2} \nu (\omega-2)^{2} (\omega-4)}.
\end{eqnarray}
In this paper, we set $ \nu = 4 $. 

\subsection{General Relations for Distributions}
The following relations are valid for any distribution. Thus, the generalized Boltzmann factor is given as:
\begin{equation}
  B(E)=\exp(-\beta_{0}E)\left(1 + \frac{1}{2} \sigma^{2} E^{2} + \sum_{r=3}^{\infty} \frac{(-1)^{r}}{r!} \langle(\beta - \beta_{0} )^{r} E^{r}\rangle\right),
\end{equation}
where
\begin{eqnarray} \nonumber
\langle(\beta - \beta_{0})^{r}\rangle &=& \sum_{j=0}^{r} \binom{r}{j} \langle\beta^{j}\rangle (-\beta_{0})^{r-j}  \\
\sigma^{2} &=& (q-1) (\beta_{0})^{2}  \\
q &=& \frac{\langle\beta^{2}\rangle}{\langle\beta\rangle^{2}}.
\end{eqnarray}
If the generalized Boltzmann factor is expressed in terms of $q$ and $\beta_{0}$, we obtain:
\begin{equation}
  B(E)= \exp(-\beta_{0} E) \left(1+ (q-1) \beta_{0}^{2} E^{2} + g(q) \beta_{0}^{3} E^{3} +\dots\right),
\end{equation}
where $ g(q) $ is a function that varies depending on the distribution. For uniform and two-level distributions, $ g(q) $ is given by:
\begin{equation}
  g(q)=0.
\end{equation}
For the gamma distribution, $ g(q) $ is:
\begin{equation}
  g(q)=- \frac{1}{3} (q-1)^{2}.
\end{equation}
Thus, for the log-normal distribution, we have:
\begin{equation}
  g(q)= - \frac{1}{6} (q^{3}- 3q +2).
\end{equation}
Moreover, for the F-distribution with $ \nu=4 $, $g(q)$ is:
\begin{equation}
  g(q)= - \frac{1}{3} \frac{(q-1) (5q-6)}{3-q}.
\end{equation}
When $q=1$, the generalized Boltzmann factor for all distributions reduces to the standard Boltzmann factor in statistical mechanics. For small fluctuations of $\beta$, the behavior of all distributions is similar, and their Boltzmann factors are comparable. However, for larger fluctuations, they can be distinguished, and the Boltzmann factors for each distribution exhibit distinct characteristics in the third and higher-order terms. In the next section, we calculate the partition function and compare the thermodynamic properties of the classical oscillator under standard and superstatistical frameworks.

\section{Thermodynamic Properties}
We aim to derive the classical oscillator's thermodynamic properties under standard and superstatistical frameworks. To do this, we first calculate the partition function and then apply the following relations to obtain the thermodynamic properties.

\subsection{Thermodynamic Relations}
Based on the above, we introduce the following functions to examine the thermodynamic properties. Thus, we have:
\begin{equation}
Z=\int_0^\infty B(E)\,dE.
\end{equation}
The above equation is the partition function. Now, we have some relations for the internal energy, Helmholtz free energy, and entropy:
\begin{equation}
U=-\frac{\partial}{\partial \beta} \ln(Z)= K_{b} T^{2} \frac{\partial \ln(Z)}{\partial T},
\end{equation}
\begin{equation}
A=-\frac{1}{\beta} \ln(Z) = - K_{b} T \ln(Z),
\end{equation}
and
\begin{equation}
S=K_{b} \beta U + K_{b} \ln(Z)=K_{b} T \frac{\partial \ln(Z)}{\partial T} + K_{b} \ln(Z).
\end{equation}

\subsection{Standard Statistical Mechanics}
By substituting the energy eigenvalues of the generalized classical oscillator into the partition function, we can derive the following thermodynamic properties:
\begin{equation}
\begin{aligned}
Z &= \frac{1}{(\hbar \omega)^N \beta^N} \\
U &= N K_B T \\
A &= K_B T N \ln(\hbar \omega) + K_B T N \ln(\beta) \\
S &= K_B N \ln(\hbar \omega) + K_B N \ln(\beta) + K_B N
\end{aligned}
\end{equation}

\subsection{Uniform and Two-Level Distributions}
Uniform and two-level distributions share the same generalized Boltzmann factor, meaning the first two terms in the generalized Boltzmann factor’s universal relation are identical. Thus, we have:
\begin{equation}
\begin{aligned}
Z &=\! \frac{1}{(\hbar \omega)^N} \left[ \frac{1}{\beta^N} + \frac{1}{2} (q-1) \frac{N(N+1)}{\beta^N} \right] \\
U &=\! K_B T N + \frac{(q-1) N(N+1)}{2} \\
A &=\! - K_B T \ln\left( \frac{1}{(\hbar \omega)^N} \left[ \frac{1}{\beta^N} + \frac{1}{2} (q-1) \frac{N(N+1)}{\beta^N} \right] \right) \\
S &=\! K_B \ln\left( Z \right) + K_B N + \frac{(q-1) N (N+1)}{2} \\
\end{aligned}
\end{equation}

Clearly, when $q=1$, all relations simplify to standard statistics.

\begin{figure}[t]
    \raggedright
    \includegraphics[width=1.1\textwidth, height=14cm]{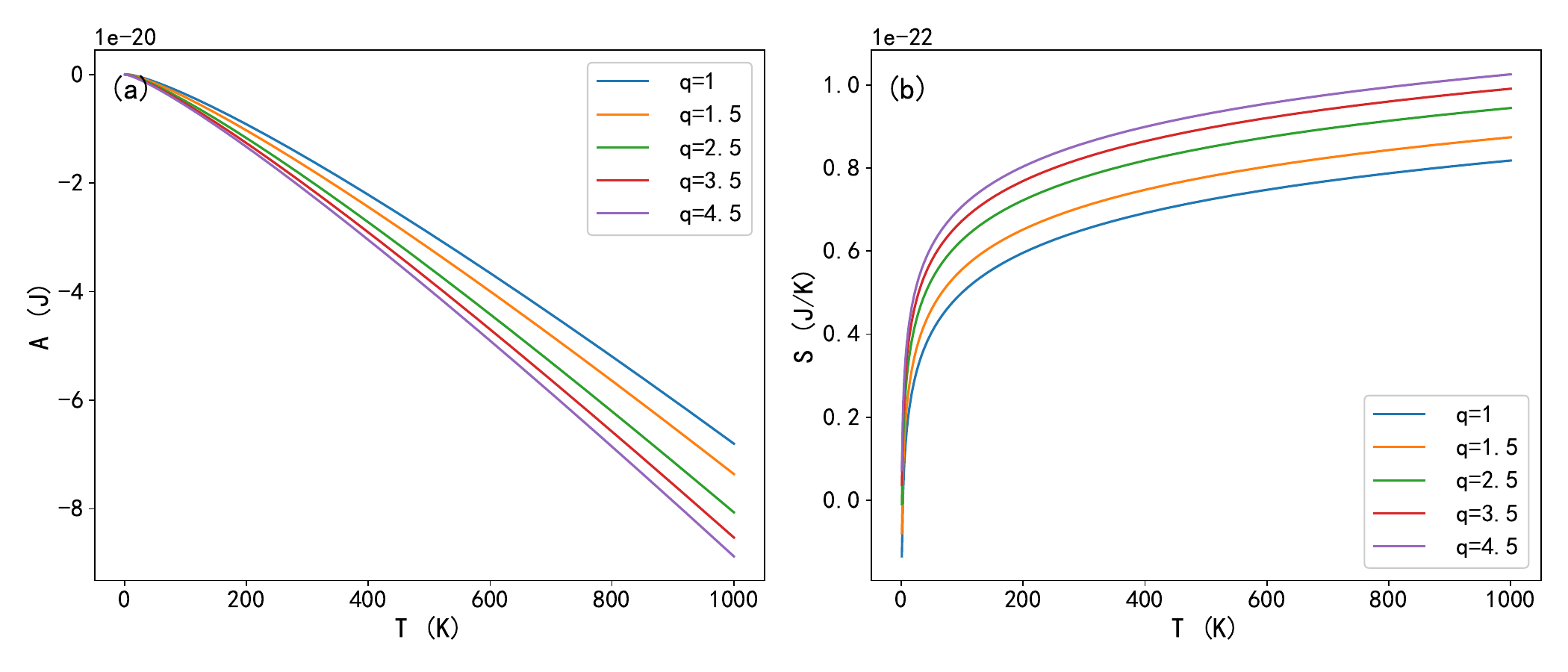} 
    \caption{Free energy (a) and entropy (b) for the uniform and 2-level distributions corresponding to \(q = 1\) (conventional statistics), 1.5, 2.5, 3.5, and 4.5.}
    \label{fig1}
\end{figure}

\subsection{Gamma Distribution}
For the gamma distribution, we obtain:
\begin{equation}
\begin{aligned}
Z &= \frac{1}{(\hbar \omega)^N} \Biggl[ \frac{1}{\beta^N} 
    + \frac{1}{2} (q-1) \frac{N(N+1)}{\beta^N}
    - \frac{1}{3} \frac{(q-1)(5q-6)}{3-q} \frac{(N+2)(N+1)}{\beta^N} \Biggr] \\
U &= K_B T\, N 
    + \frac{(q-1) N(N+1)}{2} 
    - \frac{1}{3} \frac{(q-1)(5q-6)}{3-q} (N+2)(N+1) \\
A &= - K_B T \ln\Biggl( \frac{1}{(\hbar \omega)^N} \Biggl[ \frac{1}{\beta^N} 
    + \frac{1}{2} (q-1) \frac{N(N+1)}{\beta^N} 
    - \frac{1}{3} \frac{(q-1)(5q-6)}{3-q} \frac{(N+2)(N+1)}{\beta^N} \Biggr] \Biggr) \\
S &= K_B \ln\Bigl( Z \Bigr) + K_B\, N 
    + \frac{(q-1) N (N+1)}{2} 
    - \frac{1}{3} \frac{(q-1)(5q-6)}{3-q} (N+2)(N+1)
\end{aligned}
\end{equation}

In this case, when $q=1$, these relations also simplify to standard statistics.

\begin{figure}[t]
\raggedright
\includegraphics[width=1.1\textwidth, height=14cm]{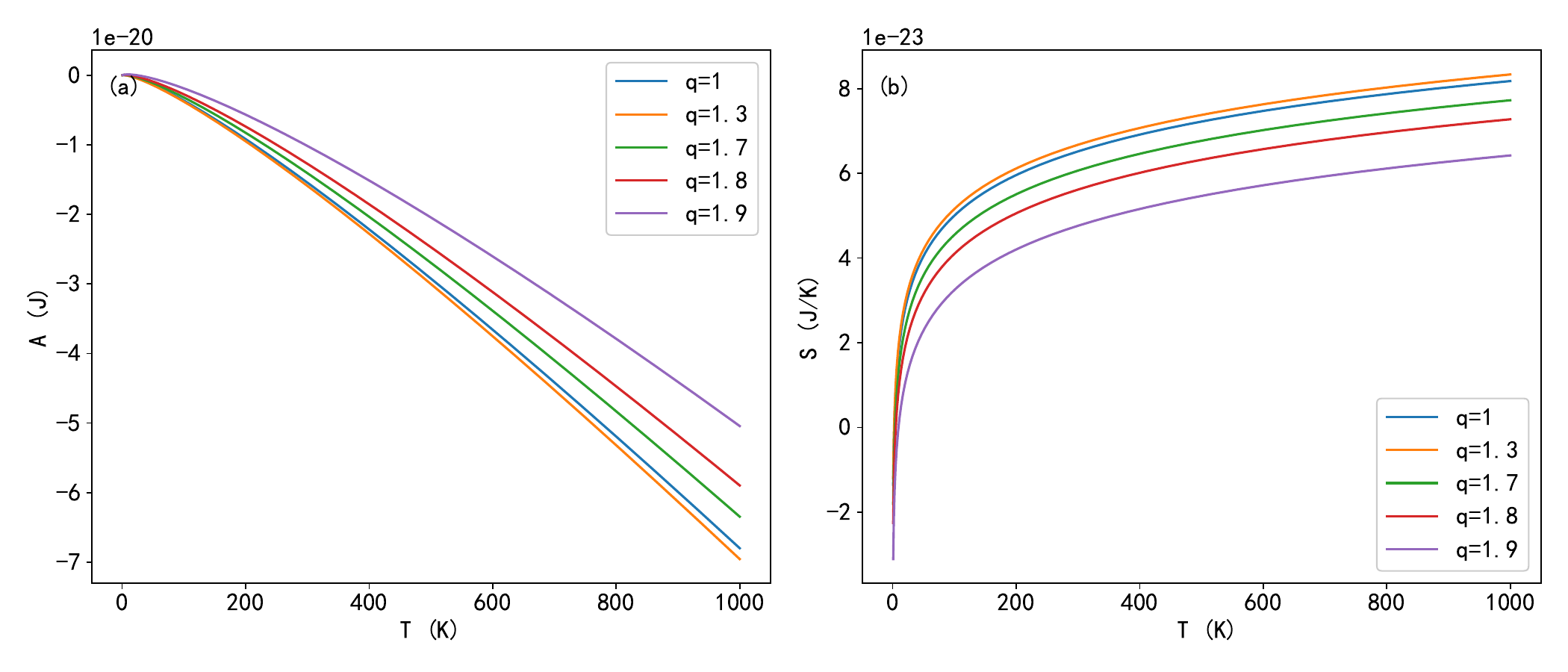}
\caption{Free energy (a) and entropy (b) for the Gamma distribution corresponding to \(q = 1,\; 1.3,\; 1.7,\; 1.8,\) and \(1.9\); note that \(q = 1.5\) also recovers conventional statistics.}\label{fig2}
\end{figure}

\subsection{Log-Normal Distribution}
The thermodynamic properties for the log-normal distribution are as follows:
\begin{equation}
\begin{aligned}
Z &= \frac{1}{(\hbar \omega)^N} \left[ \frac{1}{\beta^N} + \frac{1}{2} (q-1) \frac{N(N+1)}{\beta^N} - \frac{1}{6}(q^3 - 3q + 2) \frac{(N+2)(N+1)}{\beta^N} \right] \\
U &= K_B T N + \frac{(q-1) N(N+1)}{2} - \frac{1}{6}(q^3 - 3q + 2) (N+2)(N+1) \\
A &= - K_B T \ln\left( \frac{1}{(\hbar \omega)^N} \left[ \frac{1}{\beta^N} + \frac{1}{2} (q-1) \frac{N(N+1)}{\beta^N} - \frac{1}{6}(q^3 - 3q + 2) \frac{(N+2)(N+1)}{\beta^N} \right] \right) \\
S &= K_B \ln\left( Z \right) + K_B N + \frac{(q-1) N (N+1)}{2} - \frac{1}{6}(q^3 - 3q + 2) (N+2)(N+1) \\
\end{aligned}
\end{equation}

Similarly, for this case, when $q=1$, these relations reduce to standard statistics.

\begin{figure}[t]
\centering
\includegraphics[width=1.1\textwidth, height=14cm]{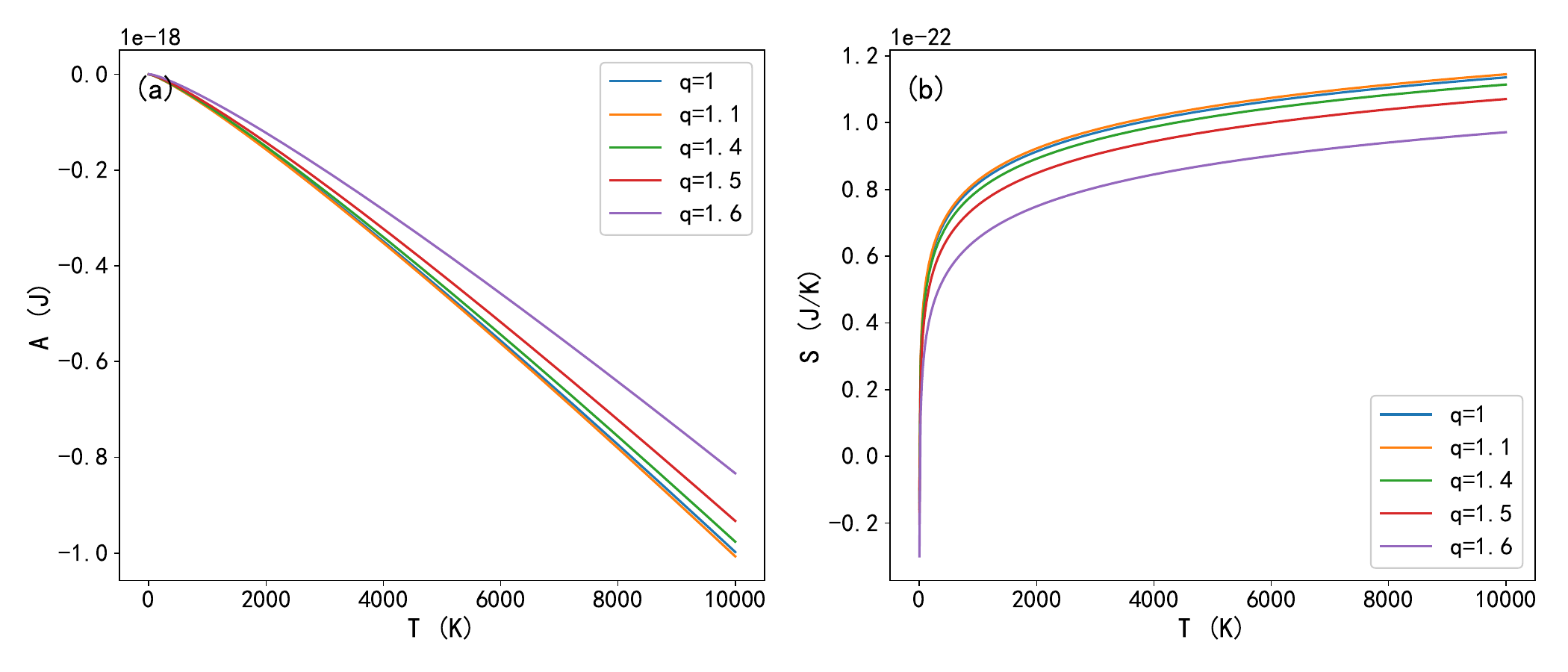}
\caption{Free energy (a) and entropy (b) for the log-normal distribution corresponding to \(q = 1,\; 1.1,\; 1.4,\; 1.5,\; 1.6\); note that \(q = 1.3\) recovers conventional statistics}\label{fig3}
\end{figure}

\subsection{F-Distribution}
As described in this paper, we assume $\nu=4$ for the F-distribution, leading to a special case. Thus, we have:
\begin{eqnarray}
\begin{aligned}
Z &= \frac{1}{(\hbar \omega)^N} \left[ \frac{1}{\beta^N} + \frac{1}{2} (q-1) \frac{N(N+1)}{\beta^N} - \frac{1}{3} \frac{(q-1)(5q-6)}{3-q} \frac{(N+2)(N+1)}{\beta^N} \right] \\
U &= K_B T N + \frac{(q-1) N(N+1)}{2} - \frac{1}{3} \frac{(q-1)(5q-6)}{3-q} (N+2)(N+1) \\
A &= - K_B T \ln\left( \frac{1}{(\hbar \omega)^N} \left[ \frac{1}{\beta^N} + \frac{1}{2} (q-1) \frac{N(N+1)}{\beta^N} - \frac{1}{3} \frac{(q-1)(5q-6)}{3-q} \frac{(N+2)(N+1)}{\beta^N} \right] \right) \\
S &= K_B \ln\left( Z \right) + K_B N + \frac{(q-1) N (N+1)}{2} - \frac{1}{3} \frac{(q-1)(5q-6)}{3-q} (N+2)(N+1) \\
\end{aligned}
\end{eqnarray}

In this special case of the F-distribution, when $q=1$, these relations do not simplify to standard statistics.
\begin{figure}[t]
\centering
\includegraphics[width=1.1\textwidth, height=14cm]{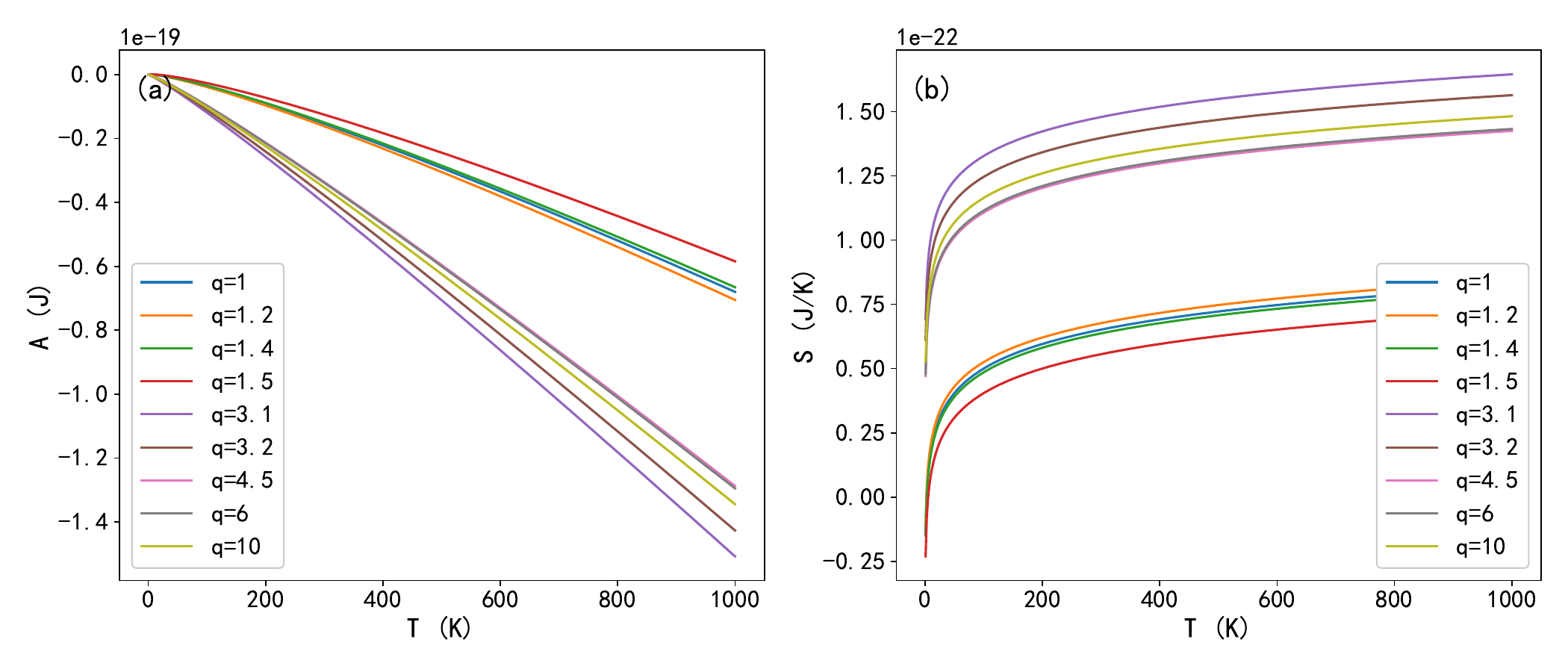}
\caption{Free energy (a) and entropy (b) for the F-distribution corresponding to \(q = 1,\; 1.2,\; 1.4,\; 1.5,\; 3.1,\; 3.2,\; 4.5,\; 6,\; 10\); note that \(q = 1.5\) also recovers conventional statistics.}\label{fig4}
\end{figure}

\section{Discussion and Results}

In this study, we investigated the changes in the thermodynamic properties of classical linear oscillators under different distributions: uniform distribution, two-level distribution, gamma distribution, log-normal distribution, and F-distribution. By analyzing the variations in Helmholtz free energy and entropy, we revealed the rules and characteristics of these distributions in nonequilibrium statistical physics and summarized their unique behaviors as the nonequilibrium parameter $q$ changes.

Firstly, for the \textbf{uniform distribution} and \textbf{two-level distribution}, their identical Boltzmann factors lead to consistent results for Helmholtz free energy and entropy. When $q = 1$, the Helmholtz free energy and entropy for both distributions revert to those of standard statistics, i.e., the system is in equilibrium. Furthermore, as $q > 1$ increases (e.g., $q = 1.5, 2.5, 3.5, 4.5$), we observe that the system's free energy gradually decreases while entropy increases. This trend indicates that as $q$ increases, the system deviates further from equilibrium, and the increase in entropy reflects a higher degree of disorder within the system. It is noteworthy that changes in $q$ can be viewed as a measure of the degree of departure from equilibrium, with larger $q$ indicating more pronounced nonequilibrium effects. Thus, uniform and two-level distributions exhibit consistent properties under nonequilibrium statistics, with trends that intuitively reflect the characteristics of $q$-statistics.

For the \textbf{gamma distribution}, the system's behavior differs from that of the uniform and two-level distributions. Under this distribution, the system follows Tsallis statistics with a $q$ range of $1 < q < 1.84$. When $q = 1$, the system reverts to standard statistics, and both Helmholtz free energy and entropy exhibit equilibrium characteristics. However, as $q$ increases, the system's free energy shows a gradual increase, while entropy decreases. This behavior contrasts sharply with that of the uniform and two-level distributions. Additionally, we find that near $q = 1.5$, the system briefly reverts to standard statistical behavior, indicating the presence of two equilibrium points: one at $q = 1$ and another at $q = 1.5$. This phenomenon implies that systems under the gamma distribution can temporarily return to equilibrium under certain conditions. However, as $q$ further increases ($q > 1.5$), the system rapidly moves away from equilibrium, exhibiting strong nonequilibrium characteristics. The existence of dual equilibrium points, along with the opposing trends in free energy and entropy during different phases, highlights the rich physical significance of the gamma distribution in nonequilibrium statistical physics.

For the \textbf{log-normal distribution}, the system's behavior shows some similarity to that of the gamma distribution, with a $q$ range of $1 < q < 1.67$. At $q = 1$, the system also reverts to standard statistics. However, as $q$ increases from 1, we observe that the system's free energy gradually increases while entropy decreases, similar to the gamma distribution. Notably, near $q = 1.3$, the system briefly returns to standard statistics, indicating two equilibrium points for this distribution as well. Unlike the gamma distribution, the changes in free energy and entropy near $q = 1.3$ are more gradual, suggesting a slower approach to equilibrium. This slow return and behavior near equilibrium may be related to the statistical properties of the log-normal distribution itself. When $q > 1.3$, the system moves away from equilibrium more quickly, eventually exhibiting pronounced nonequilibrium characteristics. In summary, the log-normal distribution demonstrates unique return-to-equilibrium and equilibrium behaviors, offering new perspectives for understanding nonequilibrium systems.

Finally, for the \textbf{F-distribution}, the system's behavior also shows similarities to the gamma and log-normal distributions, with $q$ values ranging from $1 < q < 1.58$ and $q > 3$. At $q = 1$, the system reverts to standard statistics, displaying equilibrium characteristics. As $q$ increases, the system's free energy gradually increases while entropy decreases. Near $q = 1.36$, the system again reverts to standard statistics, demonstrating two equilibrium points for this distribution. This dual equilibrium phenomenon further illustrates the complex dynamics of systems under the F-distribution during nonequilibrium evolution. Notably, when $q > 1.36$, the system moves away from equilibrium more quickly, eventually exhibiting pronounced nonequilibrium characteristics. For $q > 3$, we observe that the system's free energy continues to increase, and entropy decreases at a rapid rate. However, beyond $q > 4.8$, the system's free energy starts to decrease while entropy increases at a slower rate, exhibiting richer dynamic characteristics.

In conclusion, our study reveals the rules governing the changes in the thermodynamic properties of classical linear oscillators under different distributions as the nonequilibrium parameter $q$ varies. These distributions exhibit rich physical behaviors during their departure from equilibrium, especially under the gamma, log-normal, and F-distributions, where the phenomenon of dual equilibrium points reflects the nonequilibrium evolution characteristics of systems at different stages. These results not only deepen our understanding of nonequilibrium statistical physics but also provide theoretical foundations and new research directions for studying nonequilibrium behaviors in complex systems.

\section{Conclusion}

This study explored the changes in the thermodynamic properties of classical linear oscillators under various statistical distributions as the nonequilibrium parameter $q$ varies. For uniform and two-level distributions, the system exhibits consistent behavior under nonequilibrium statistics, with free energy decreasing and entropy increasing as $q > 1$, indicating a gradual departure from equilibrium. The gamma, log-normal, and F-distributions exhibit more complex dynamic characteristics, particularly with the presence of dual equilibrium points at specific $q$ values, allowing the system to temporarily return to equilibrium under certain conditions. This phenomenon highlights the rich physical significance of these distributions in nonequilibrium evolution, reflecting the unique characteristics of systems at different stages. Furthermore, the study found that as $q$ increases, the speed and manner of departure from equilibrium vary across distributions. Notably, in the F-distribution, when $q > 3$, the trends in free energy and entropy reverse again, demonstrating even more complex dynamic behavior. These results not only expand the theoretical framework of nonequilibrium statistical physics but also provide new perspectives for understanding and describing the behavior of complex systems under nonequilibrium conditions.

Future research could explore the effects of other types of statistical distributions on the thermodynamic properties of systems and validate the theoretical predictions with experimental data. Additionally, considering the interactions between systems and the impact of higher dimensions could help comprehensively understand the evolutionary laws of complex systems under nonequilibrium conditions.

\textbf{ACKNOWLEDGMENTS:} 
This work was supported in part by the Fundamental Research Funds for the Central Universities, China (Grant No. CCNU19QN029), and the National Natural Science Foundation of China (Grant No. 61873104).

\section{Data Availability Statement}
No new data were created or analysed in this study as it is a theoretical work.



\section{Appendix A: Illustrations of Different Distributions}

In this section, we plot the Helmholtz free energy and entropy as functions of temperature, analyzing them for different values of the Tsallis parameter and distributions. The Helmholtz free energy is a thermodynamic potential that represents the maximum amount of work a system can perform under isothermal conditions. Entropy, on the other hand, is a measure of the disorder or randomness of a system, reflecting the number of possible microstates.

In this study, we use a consistent set of fixed parameters across all the plots. The graphs include curves of the Helmholtz free energy and entropy as functions of temperature, where the temperature is measured in Kelvin. In each graph, curves related to ordinary statistics and superstatistical distributions are compared for different values of the Tsallis parameter \( q \) (e.g., \( q=1.0, 1.5, 2.0 \), etc.).

The relevant parameters and their units are listed below:

\begin{itemize}
    \item Boltzmann constant: \( K_b = 1.3806488 \times 10^{-23} \, \mathrm{J/K} \)
    \item Planck constant: \( \hbar = 1.053472010 \times 10^{-34} \, \mathrm{J \cdot s} \)
    \item Angular frequency: \( \omega = 0.95 \times 10^{12} \, \mathrm{rad/s} \)
    \item particle number:$N=1$
\end{itemize}




\begin{thebibliography}{10}
\expandafter\ifx\csname url\endcsname\relax
  \def\url#1{\texttt{#1}}\fi
\expandafter\ifx\csname urlprefix\endcsname\relax\def\urlprefix{URL }\fi
\expandafter\ifx\csname href\endcsname\relax
  \def\href#1#2{#2} \def\path#1{#1}\fi

\bibitem{gitterman2005classical}
M.~Gitterman, Classical harmonic oscillator with multiplicative noise, Physica A: Statistical Mechanics and its Applications 352~(2-4) (2005) 309--334.

\bibitem{jimenez2008brownian}
J.~Jim{\'e}nez-Aquino, R.~Velasco, F.~Uribe, Brownian motion of a classical harmonic oscillator in a magnetic field, Physical Review E—Statistical, Nonlinear, and Soft Matter Physics 77~(5) (2008) 051105.

\bibitem{giannakis2021quantum}
D.~Giannakis, Quantum dynamics of the classical harmonic oscillator, Journal of Mathematical Physics 62~(4) (2021).

\bibitem{dekker1981classical}
H.~Dekker, Classical and quantum mechanics of the damped harmonic oscillator, Physics Reports 80~(1) (1981) 1--110.

\bibitem{bloch2013introduction}
S.~C. Bloch, Introduction to classical and quantum harmonic oscillators, John Wiley \& Sons, 2013.

\bibitem{li2025multi}
J.~Li, H.~Gurgenci, Z.~Guan, J.~Wang, J.~Chen, C.~Wang, Z.~Huang, Multi objective optimization algorithm for hybrid quantum harmonic oscillator and its application in rotor system optimization, Scientific Reports 15~(1) (2025) 7534.

\bibitem{onah2023quadratic}
F.~Onah, E.~Garc{\'\i}a~Herrera, J.~Ruelas-Galv{\'a}n, G.~Ju{\'a}rez~Rangel, E.~Real~Norzagaray, B.~Rodr{\'\i}guez-Lara, A quadratic time-dependent quantum harmonic oscillator, Scientific Reports 13~(1) (2023) 8312.

\bibitem{alves2023algebraic}
M.~B. Alves, Algebraic solution for the classical harmonic oscillator, Revista Brasileira de Ensino de F{\'\i}sica 45 (2023) e20230152.

\bibitem{jimenez2009fluctuation}
J.~Jim{\'e}nez-Aquino, R.~Velasco, F.~Uribe, Fluctuation relations for a classical harmonic oscillator in an electromagnetic field, Physical Review E—Statistical, Nonlinear, and Soft Matter Physics 79~(6) (2009) 061109.

\bibitem{huang2015discrete}
W.~C.-W. Huang, H.~Batelaan, Discrete excitation spectrum of a classical harmonic oscillator in zero-point radiation, Foundations of Physics 45 (2015) 333--353.

\bibitem{sattin2018superstatistics}
F.~Sattin, Superstatistics and temperature fluctuations, Physics Letters A 382~(36) (2018) 2551--2554.

\bibitem{davis2025superstatistics}
S.~Davis, C.~Loyola, C.~Femen{\'\i}as, J.~Peralta, Superstatistics as the thermodynamic limit of driven classical systems, Physica A: Statistical Mechanics and its Applications (2025) 130370.

\bibitem{metzler2020superstatistics}
R.~Metzler, Superstatistics and non-gaussian diffusion, The European Physical Journal Special Topics 229~(5) (2020) 711--728.

\bibitem{ourabah2024superstatistics}
K.~Ourabah, Superstatistics from a dynamical perspective: Entropy and relaxation, Physical Review E 109~(1) (2024) 014127.

\bibitem{dos2020log}
M.~A.~F. dos Santos, L.~Menon~Junior, Log-normal superstatistics for brownian particles in a heterogeneous environment, Physics 2~(4) (2020) 571--586.

\bibitem{beck2003superstatistics}
C.~Beck, E.~G. Cohen, Superstatistics, Physica A: Statistical mechanics and its applications 322 (2003) 267--275.

\bibitem{beck2004superstatistics}
C.~Beck, Superstatistics: theory and applications, Continuum mechanics and thermodynamics 16 (2004) 293--304.

\bibitem{cohen2004superstatistics}
E.~Cohen, Superstatistics, Physica D: Nonlinear Phenomena 193~(1-4) (2004) 35--52.

\bibitem{abe2007superstatistics}
S.~Abe, C.~Beck, E.~G. Cohen, Superstatistics, thermodynamics, and fluctuations, Physical Review E—Statistical, Nonlinear, and Soft Matter Physics 76~(3) (2007) 031102.

\bibitem{beck2009recent}
C.~Beck, Recent developments in superstatistics, Brazilian Journal of Physics 39 (2009) 357--363.

\bibitem{gravanis2020physical}
E.~Gravanis, E.~Akylas, G.~Livadiotis, Physical meaning of temperature in superstatistics, Europhysics Letters 130~(3) (2020) 30005.

\bibitem{dos2023entropic}
M.~A. Dos~Santos, F.~D. Nobre, E.~M. Curado, Entropic form emergent from superstatistics, Physical Review E 107~(1) (2023) 014132.

\bibitem{sattin2004superstatistics}
F.~Sattin, Superstatistics from a different perspective, Physica A: Statistical Mechanics and its Applications 338~(3-4) (2004) 437--444.

\bibitem{souza2003stability}
A.~Souza, C.~Tsallis, Stability of the entropy for superstatistics, Physics Letters A 319~(3-4) (2003) 273--278.

\bibitem{sadeghi2024investigation}
J.~Sadeghi, S.~N. Gashti, et~al., Investigation of thermal properties of hulth{\'e}n potential from statistical and superstatistical perspectives with various distributions, Physica Scripta 99~(9) (2024) 095254.

\end{thebibliography}



\end{document}